# Anomalous Cosmic Ray Oxygen Observations in to 0.1 au

J. S. Rankin[1], D. J. McComas[1], R. A. Leske[2], E. R. Christian[3], C. M. S. Cohen[2], A. C. Cummings[2], C. J. Joyce[1], A. W. Labrador[2], R. A. Mewaldt[2], N. A. Schwadron[1,4], E. C. Stone[2], R. D. Strauss[5], M. E. Wiedenbeck[6]

1 Department of Astrophysical Sciences, Princeton University, Princeton, NJ 08540, USA
2 California Institute of Technology, Pasadena, CA 91125, USA
3 NASA Goddard Space Flight Center, Greenbelt, MD 20771, USA
4 University of New Hampshire, Durham, NH 03824, USA
5 Center for Space Research, North-West University, Potchefstroom, South Africa
6 Jet Propulsion Laboratory, California Institute of Technology, Pasadena, CA 91109, USA

## Abstract

The Integrated Science Investigation of the Sun instrument suite onboard NASA's *Parker Solar Probe* mission continues to measure solar energetic particles and cosmic rays closer to the Sun than ever before. Here, we present the first observations of cosmic rays into 0.1 au (21.5 solar radii), focusing specifically on oxygen from ~2018.7 to ~2021.2. Our energy spectra reveal an anomalous cosmic ray-dominated profile that is comparable to that at 1 au, across multiple solar cycle minima. The galactic cosmic ray-dominated component is similar to that of the previous solar minimum (Solar Cycle 24/25 compared to 23/24) but elevated compared to the past (Solar Cycle 20/21). The findings are generally consistent with the current trend of unusually weak solar modulation that originated during the previous solar minimum and continues today. We also find a strong radial intensity gradient: $49.4 \pm 8.0$ % au$^{-1}$ from 0.1 to 0.94 au, for energies of 6.9 to 27 MeV nuc$^{-1}$. This value agrees with that measured by Helios nearly 45 years ago from 0.3 to 1.0 au ($48 \pm 12$ % au$^{-1}$; 9 to 29 MeV nuc$^{-1}$) and is larger than predicted by models. The large ACR gradients observed close to the Sun by the Parker Solar Probe Integrated Science Investigation of the Sun instrument suite found here suggest that intermediate-scale variations in the magnetic field's structure strongly influences cosmic ray drifts, well inside 1 au.

## 1. Introduction

The study of the spatial and temporal variations of anomalous cosmic rays (ACRs) has been ongoing for nearly 50 years and continues to provide compelling insight about the nature of our heliosphere and the acceleration and transport of energetic particles. ACRs consist mainly of singly-ionized atomic species of high first-ionization potential elements – such as hydrogen, helium, nitrogen, oxygen, neon, and argon – typically in the ~5 to ~50 MeV nuc$^{-1}$ energy range (e.g., Garcia-Munoz et al. 1973; Hovestadt et al. 1973; McDonald et al. 1974; Christian et al. 1988; Klecker et al. 1998; Cummings et al. 2002a, 2002b; Potgieter 2013). Originating as interstellar neutrals, these particles drift into the heliosphere as a part of the interstellar wind at speeds of ~25.4 km s$^{-1}$ (McComas et al. 2015). On their journey inward toward the Sun, some fraction gets ionized (Fisk et al. 1974), forming "pickup ions" which gain energy in the motional electric field as they are carried back outward as a part of the solar wind (e.g., Möbius et al. 1985; Drews et al. 2016). A tiny portion of these ~keV pickup ions are accelerated in the outer heliosphere, where they gain energies of a few to a hundred MeV nuc$^{-1}$ over timescales of a year or less (Jokipii 1996; Mewaldt et al. 1996; Barghouty et al. 2000; Giacalone et al. 2012). The newly accelerated particles form a new population at the low-energy end of the cosmic ray spectra, known as "anomalous cosmic rays". The source and mechanism for ACR acceleration is still a matter of debate, as reflected in multiple theories, including: compressive turbulence in the heliosheath (Fisk & Gloeckler 2009), magnetic reconnection near the heliopause (Drake et al. 2010), second-order Fermi processes (Strauss et al. 2010), and acceleration to higher energies preferentially towards the flanks and tail of the blunt termination shock (McComas & Schwadron 2006). Of these, the latter is the best supported by current observations and used by many models (McComas & Schwadron 2006; Kóta & Jokipii 2008; Schwadron et al. 2008; Guo et al. 2010; Kóta 2010; Senanayake & Florinski 2013; McComas et al. 2019a).

In general, the transport of cosmic rays through the heliosphere involves a complex interplay of different physical processes such as: (i) diffusion caused by irregularities in the magnetic field, (ii) adiabatic energy loss, (iii) convection in the outwardly expanding solar wind, and (iv) gradient and curvature drifts in the large-scale heliospheric magnetic field. Parker (1965) captured these effects remarkably well in the widely-used Parker transport equation (e.g., Jokipii & Levy 1977; Jokipii & Thomas 1981). However, the relationship amongst these parameters, how their spatial and temporal variations influence the large-scale structure of the heliosphere, and, for example, the applicability of the transport equation near the Sun, are not yet fully understood (see, e.g., discussions by Fujii & McDonald 1997; Fisk et al. 1998; Rankin et al. 2021).

The spatial distribution of cosmic rays as a function of latitude and radial distance has been studied by several spacecraft on their specific passages through the heliosphere (see, e.g. Webber et al. 1975, 1981; Marsden et al. 1999; Fuji et al. 1999; Cummings et al. 1990, 1995, 2009). This distribution is strongly influenced by a complex interplay of drift and diffusion – the configuration of the mean magnetic field and ability of particles to move along and across it. For example, radial intensity gradients are weaker in the outer heliosphere compared to the inner heliosphere (see, e.g. Cummings et al. 1987). To first order, this is attributed to the nature of the diffusion coefficient, which is often assumed to be inversely proportional to the heliospheric magnetic field. The Parker Spiral field geometry comprises a radial component of the field that drops off as $r^{-2}$ and a tangential component that goes as $r^{-1}$. The field smoothly transitions from radial to transverse with distance (~45-deg angle at 1 au and nearly ~90-deg by 10 au), causing the field strength to also vary with distance (decreasing as ~$r^{-2}$ near the Sun and transitioning closer to ~$r^{-1}$ within several au, in the ecliptic). However, the combination of fast and slow wind, as well as turbulence complicates this simple picture, as has been noted by many authors (e.g., Engelbrecht et al. 2017; Zhao et al. 2018; Shen et al. 2019; Moloto & Engelbrecht 2020 and references therein).

In addition to spatial variations, cosmic ray energy spectra are observed to undergo cyclical temporal changes in intensity that are anticorrelated with the solar cycle. This process, known as "solar modulation", follows both the 11-year pattern of solar activity as well as the longer-term 22-year polarity cycle. During negative polarity cycles (qA < 0; e.g., the SC 23/24 minimum), positively charged ions generally drift inward along the heliographic equator and outward through the north and south polar regions, whereas during positive polarity cycles (qA > 0; e.g., the SC 24/25 minimum), the opposite occurs: particles drift inward from the poles and outward along the equator. Since drifts occur via a charge-sign-dependent process, the behavior is reversed for negatively charged particles such as antiprotons and electrons (Jokipii et al. 1977; Potgieter 1998; Potgieter 2013; Potgieter 2017). These phase-dependent global drift patterns produce several notable spatial effects, including: (i) radial gradients that are stronger and more responsive to the tilt of the heliospheric current sheet during qA < 0 as opposed to qA > 0 cycles (see, e.g. McDonald 1998; Cummings & Stone 1999; Stone & Cummings 1999; Cummings et al.

1995, 2009, and references therein), (ii) sign-dependent latitudinal gradients that are, for the most part[1], positive during qA > 0 cycles and negative during qA < 0 cycles (McKibben 1987, 1989; Cummings et al. 1987, 1995; Marsden et al. 1999; Cummings et al. 2009; Ngobeni & Potgieter 2010), and (iii) a likely polarity-dependent source location for ACR acceleration, owing to maximum intensities at the termination shock near the equator during qA < 0 cycles and towards the poles during qA > 0 periods (see, e.g., Fisk et al. 1998; Cummings & Stone 1999, and references therein).

From the time of the discovery of ACRs (late 1970's) until 2009, solar modulation patterns for ACRs and GCRs showed persistently similar behavior. However, an unexpected deviation from this trend occurred in 2009, during the unusually long Solar Cycle (SC) 23/24 solar minimum. While the GCRs achieved record-high intensities (~20 to 26% above previous levels; see, e.g., Mewaldt et al. 2010), ACRs reached only ~90% of their mid-1997 levels before declining abruptly in early 2010 (Leske et al. 2013; see also McDonald et al. 2010). The SC 23/24 solar minimum exhibited other unusual features as well, including a reduced solar wind dynamic pressure and heliospheric magnetic field (McComas et al. 2008), a less turbulent magnetic field (Smith & Balogh 2008), and a slowly declining heliospheric current sheet tilt angle (Wang et al. 2009). Together, these observations could provide a natural explanation for the record-setting intensities of GCRs. For example, Mewaldt et al. (2010) estimated that the parallel diffusion coefficient was ~44% greater in 2009 (SC 23/24) than in the previous 1997-1998 minimum (SC 22/23). The resulting longer mean free paths translated to less solar modulation, leading to increased intensities at 1 au (see also Strauss & Potgieter 2014).

A larger diffusion coefficient may also explain the ACR-GCR discrepancy. According to Moraal & Stoker (2010), ACRs, like GCRs, likely underwent less modulation compared to previous solar cycles, however, they also experienced a lowered acceleration efficiency at the termination shock. The recent cycle 24/25 minimum follows what was the weakest-magnitude cycle of the space age (see, e.g., Hajra 2021) and the trend of GCR-ACR discrepancy continues. The slightly weaker magnetic field and low levels of turbulence have led to an estimated ~10% increase of the cosmic ray mean free path compared to 2009, leading to a new record high for GCR intensities while ACRs levels remain similar to those of previous epochs (see, e.g. Fu et al. 2021). However, there have also been some notable differences between this minimum and the previous one. For instance, the heliospheric current sheet achieved a minimum of ~2.1-deg in April 2020 that was 22% lower than 22/23 and 53% lower than 23/24. Further, the CME eruption rate was less than half that during solar minima 22/23 and 23/24. These differences, combined with opposite drift patterns makes this a compelling time to compare the behavior of ACRs to GCRs and put basic transport theory and possible source location of ACRs to the test by studying spatial gradients near the Sun.

Launched August 12, 2018, NASA's *Parker Solar Probe* (*PSP*) spacecraft (Fox et al. 2016), continues to make surprising discoveries in previously unexplored regions near the Sun, inside 0.25 au. PSP achieves progressively lower perihelion distances every few solar encounters – using a series of Venus flybys to reduce angular momentum (Guo et al. 2021) – and is expected to attain an ultimate orbit reaching under 10 solar radii (< 0.045 au) by the end of 2024. Measurements so far have taken place during the minimum between solar cycles 24 and 25. The low levels of solar activity during this time have provided a prime opportunity for the Integrated Science Investigation of the Sun (ISØIS) instrument suite (McComas et al. 2016) to observe cosmic rays closer to the Sun than ever before (McComas et al. 2019b). Rankin et al. (2021) recently reported on the first observations of ACRs in to 36 solar radii (0.166 au). These authors examined PSP/ISOIS measurements from ~2018.7 to 2019.9 and, after subtracting out GCRs, found radial gradients of $34.3 \pm 5.6$ % au$^{-1}$ and $44.7 \pm 10.2$ % au$^{-1}$ for Helium in the 4.0 to 32 MeV nuc$^{-1}$ and 13 to 45 MeV nuc$^{-1}$ energy ranges, respectively. These values were much larger than those described by prior studies of Helium in the inner heliosphere (e.g. Webber et al. 1981; Bastian et al. 1981; McKibben 1989; Cummings et al. 1990; McDonald et al. 2001; and references therein).

---

[1] The model of Strauss & Potgieter (2010) showed that, during qA < 0 phases, the latitudinal gradient can be either positive or negative depending on which of the competing terms dominate between drift and diffusion.

Here, we report the first measurements of cosmic rays in to 0.1 au, focusing specifically on oxygen. We examine the spectra observed by PSP from ~1.5 to ~100 MeV nuc$^{-1}$ and compare to those at 1 au. We also determine the magnitude of the ACR radial intensity gradient (~7 to ~27 MeV nuc$^{-1}$) and compare it to results obtained from Helios from 1975 to 1977 (Marquardt et al. 2018; 0.3 to 1.0 au). Finally, we compare these observations to those from previous solar minima and discuss their implications in the context of current understanding of the ACR source location, cosmic ray transport and solar modulation in the inner heliosphere.

## 2. Observations & Methods

This study uses measurements taken by ISOIS' High-energy Energetic Particle Instrument (EPI-Hi; McComas et al. 2016; Wiedenbeck et al. 2017) over PSP's first seven orbits (August 29, 2018, through March 2, 2021). ISOIS/EPI-Hi's three cylindrically stacked solid-state detector telescopes (LET1, LET2, and HET) span energies broad enough to include both ACR and GCR oxygen, as shown by the spectrum in Figure 1.

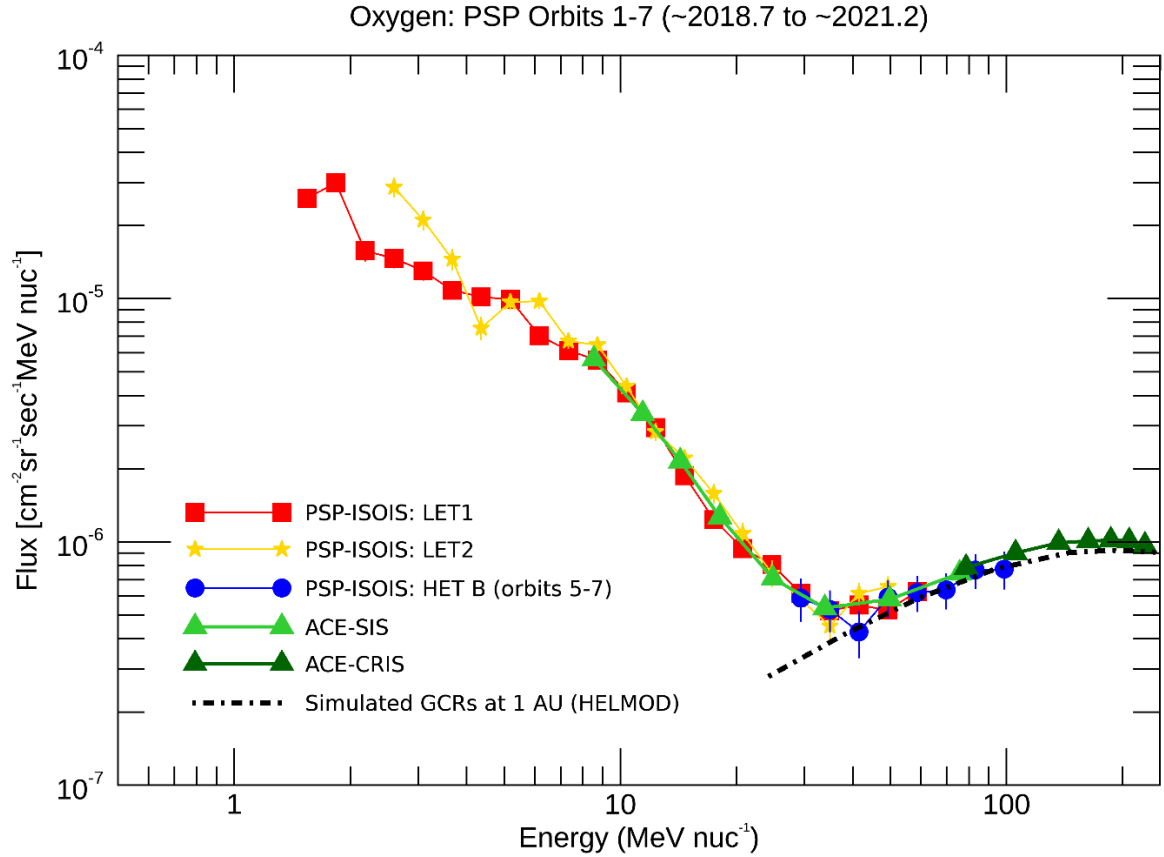

Figure 1. Event-subtracted oxygen spectra averaged over the ~2018.7 to ~2021.2 timeframe for the three EPI-Hi telescopes on PSP/ISOIS: energies LET1 (red; 1.5 to 59 MeV nuc⁻¹), LET2 (gold; 2.6 to 49 MeV nuc⁻¹), and HET (blue; 29 to 99 MeV nuc⁻¹). PSP/ISOIS observations averaged over these seven orbits agree well with 1-au observational counterparts ACE/SIS (bright green; 8.5 to 76 MeV nuc⁻¹) and ACE/CRIS (dark green; 79 to 230 MeV nuc⁻¹) and simulated GCR oxygen at 1 au derived from solar wind conditions over the same time period (dashed black lines; HelMod Online Calculator: version 4.2.1, February 2021; [www.helmod.org](http://www.helmod.org)). Due to an instrumental anomaly that occurred during orbits 1-4, the HET best captured using its B-side data taken during orbits 5-7 (HET's quiet-time A-side rates are partially contaminated with background arising from partial obstruction by PSP's thermal protection system). Data shown include 1-sigma statistical uncertainties (which, in most cases, are smaller than the symbols).

For comparison to observations at 1 au, we use data from the Solar Isotope Spectrometer (SIS; Stone et al. 1998b) and the Cosmic Ray Isotrope Spectrometer (CRIS; Stone et al. 1998c) onboard NASA's Advanced Composition Explorer (ACE; Stone et al. 1998a). As indicated by Figure 1, the ACE spectra closely agree with PSP/ISOIS. Since none of these instruments are designed to distinguish amongst charge states, the measured spectra represent a superposition of ACR and GCR populations. Therefore, we also superimpose a simulated GCR oxygen spectrum at 1 au, calculated using the HELMOD model (version 4.2.1, 2021 February; Bobik et al. 2012; Boschini et al. 2020). Since the simulated result is clearly consistent with the high-energy observations, this enables us to measure ACRs by subtracting out the GCR contribution in the transition region (e.g., the ~25 to 50 MeV nuc⁻¹ range).

To characterize the radial gradient, we use the approach developed by Rankin et al. (2021) and: i) subtract out energetic particle events, ii) correct for time-varying levels of modulation, and iii) apply log fits of the radial gradient function to the data in linear space, assuming the form:

$$g_r = \frac{1}{f}\frac{\partial f}{\partial r} = \frac{\partial \ln f}{\partial r} \quad (1)$$

for flux, $f$, radial intensity gradient, $g_r$, and radial distance, $r$ (see, e.g., Jokipii 1971; Jokipii et al. 1977; Strauss & Potgieter 2010).

The PSP orbit 1-7 time period (~2018.7 to ~2021.2) is very quiet with only two oxygen events identified in by our simple algorithm (adapted from Rankin et al. 2021). These events coincided with a coronal mass ejection that was ongoing at the time of instrument turn-on (Rankin et al. 2021; DOY 241-260 of 2018) and the mid-sized solar energetic particle event that occurred on November 29, 2020 (DOY 334-340 of 2020) described by Cohen et al. (2021). Since oxygen is much more statistically limited than helium and the LET1 and LET2 spectra are nearly identical, we combine fluxes from the three telescope apertures (LET1 A & B, LET2 C) and de-trend using ACE-SIS and ACE-CRIS data as our baseline (27-day averaged; interpolated to EPI-Hi's daily averages). Once the data has been event-subtracted and de-trended, we isolate the ACR component by first averaging over the ACR-dominated energy range (6.7 to 27 MeV nuc$^{-1}$; see Figure 1) and then subtracting the GCR component (using the simulated spectrum at 1 au; HELMOD). Owing to an instrumental anomaly that impacted HET oxygen during orbits 1-4, we omit HET from the present radial gradient analysis as the remaining data is statistically limited; therefore, a GCR oxygen radial gradient cannot be determined at this time. We fit the data binned in 0.025 au radial increments to ensure that there are 30 or more counts for each bin, enabling Gaussian statistical methods to be used in the treatment of uncertainties. Our fit results are shown in Figure 2.

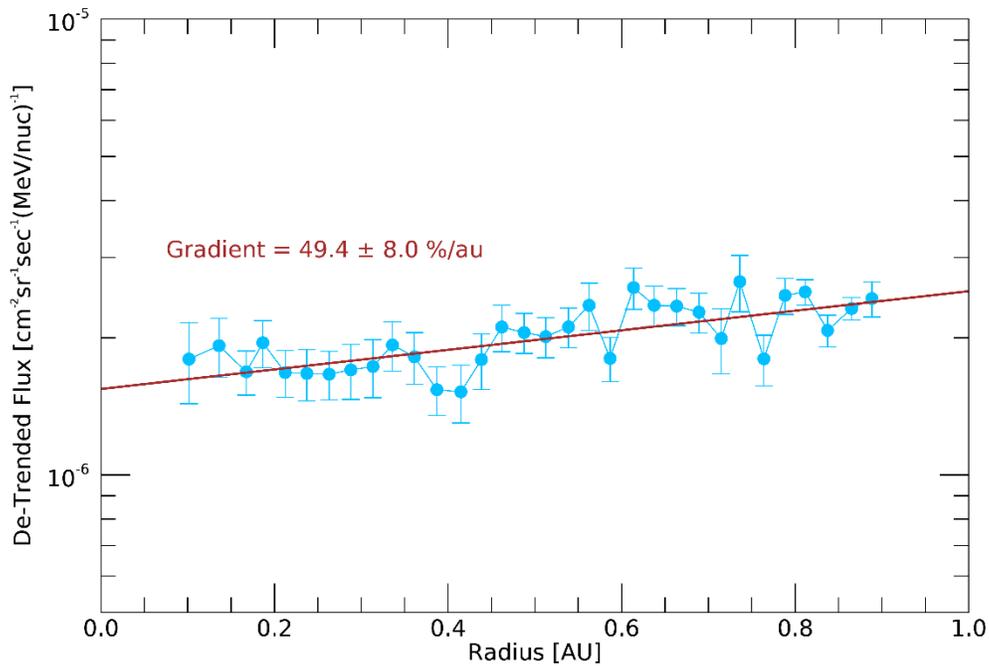

Figure 2. Radial intensity gradient determined by fitting flux as a function of radius for 6.7 to 27 MeV nuc$^{-1}$ oxygen. The flux was obtained by averaging three ISⓄIS/EPI-Hi telescope apertures (LET1 A & B and LET2 C), subtracting out energetic particle events, de-trending using ACE/SIS & CRIS as a 1-au baseline, and subtracting out the GCR component, and averaging in 0.025-au radial bins. The data were accumulated from ~2018.7 to ~2021.2.

## 3. Results & Discussion

As demonstrated in Figure 1, the ACR and GCR oxygen spectra observed by PSP/ISⓄIS (averaged over radial distances from 0.096 to 0.94 au) are very similar to that of ACE at 1 au. Perhaps even more interesting is that the ACR component of the spectra measured during the SC 24/25 minimum (PSP & ACE, ~2018.7 to 2021.2) agrees remarkably well with that measured at 1 au during the SC 20/21 minimum (IMP 8, ~1974 to 1976), around the time that Helios began taking its first measurements, as well as that of ACE (2009) during the previous solar minimum (SC 23/24), depicted in Figure 3. The results are consistent with the findings of Leske et al. (2013) who reported that, during the minimum of SC 23/24, ACR intensities at 1 au retained levels comparable to those of the preceding three minima, despite GCR intensities reaching record highs (see also Mewaldt et al. 2010). This ACR-GCR discrepancy was also noted by Moraal & Stoker (2010), who pointed out that the increase of the diffusion coefficient (resulting from a weaker heliospheric magnetic field and lower levels of turbulence) would naturally produce increased GCR intensities while having a different effect on ACRs as it would lead to a lower acceleration efficiency at the termination shock. However, why the GCRs and ACRs were so clearly correlated with each other up until then and why the ACR fluxes at 1 au produce similar spectra across multiple minima is not entirely clear. The ACR-GCR discrepancy also persists for the current cycle 24/25 minimum.

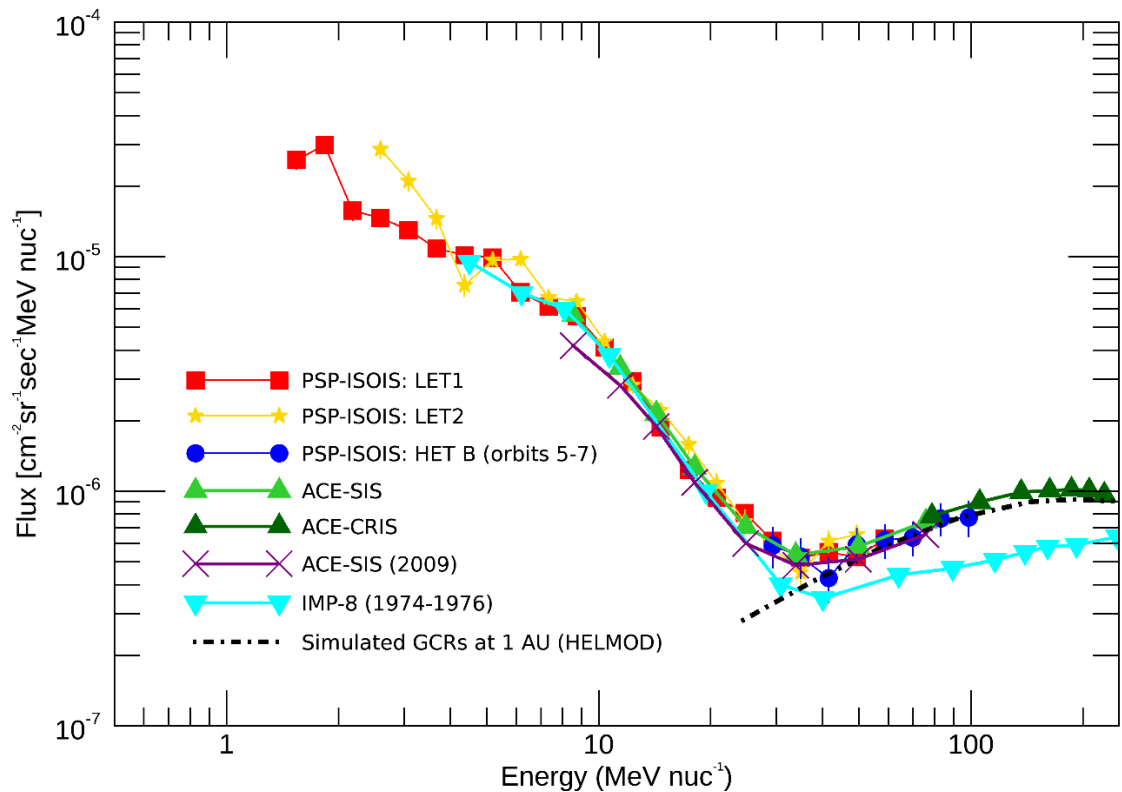

Figure 3. Quiet-time oxygen spectra over multiple solar minima. Similar to Figure 1, but with 1-au observations from the Helios era (IMP-8; cyan) and the previous solar minimum (ACE-SIS; purple) superimposed.

Figure 4 (see also Table 1 in the Appendix) compares the magnitude of the ACR oxygen radial gradient obtained above to those from previous studies in the heliosphere during the past 5 solar minima. Also included is a recent result of Marquardt et al. (2018) who re-visited Helios data and derived the magnitudes of ACR oxygen radial gradients near the Sun, from 0.3 to 1 au. Our result from PSP ($49.4 \pm 8.0$ % au$^{-1}$) agrees remarkably well with that measured by Helios ~45 years ago ($48 \pm 12$ % au$^{-1}$) for a comparable energy range (6.7 to 27 MeV nuc$^{-1}$ and 9.0 to 29 MeV nuc$^{-1}$, respectively). Beyond 1 au, there is a clear trend of decreasing gradient magnitude as a function of radial distance. Studies during negative polarity cycles ($qA < 0$ minima) indicate that radial gradients from 1 to 3 au are a factor of two larger than their positive-polarity counterparts (Cummings et al. 1990, Cummings et al. 2009). During the cycle 20/21, 23/24, and 24/25 minima, the fluxes (Figure 3; IMP-8 1974-1976, ACE-SIS 2009, and PSP-ISOIS, respectively) as well as the magnitudes of the gradients (Figure 4) are virtually indistinguishable.

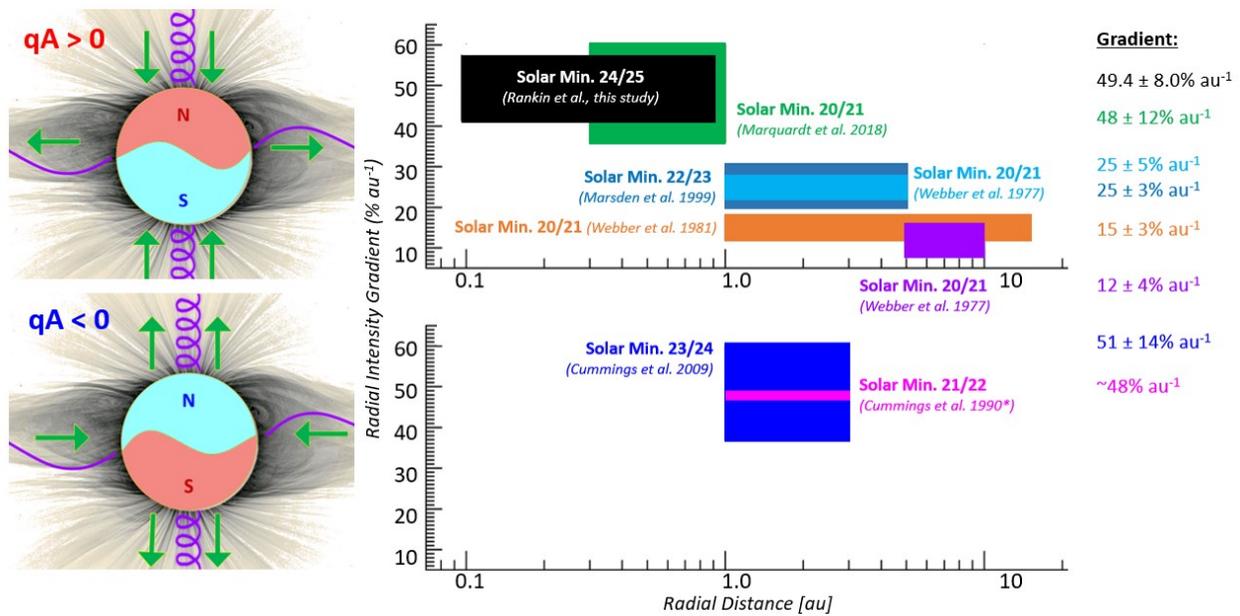

Figure 4. Comparison of ACR Oxygen radial gradients in the inner heliosphere over multiple solar minima. Right: The top panel denotes values measured during qA > 0 cycles, while the bottom panel denotes those measured during qA < 0 cycles. We note that PSP's mean radial distance is ~0.63 au because of its highly eccentric orbit. Left: Each polarity is characterized by a different global cosmic ray drift pattern, as illustrated with respect to the Sun's North (N) and South (S) magnetic fields, and its heliospheric current sheet (wavy line). The green arrows indicate drift directions for positively-charged ions.

The magnitude of the ACR oxygen radial intensity gradient observed near the Sun by both Helios and PSP differ from predictions made by models, as shown in Figure 5. Here, the modeled radial gradient, from Strauss & Potgieter (2010), is compared to a number of observed values as a function of both radius and energy, and for both drift cycles. Although the model compares rather well with observations beyond ~2 au, the modeled and observed values diverge strongly inside ~2 au; while the modeled results decrease towards the Sun, the observations seem to continue increasing as ~r⁻¹ (the blue dotted line). The reason for this discrepancy remains unsolved, but might be due to a number of reasons, including that the model is simply unable to capture the (possibly) large drift effects close to the Sun due to the assumed numerical boundary condition assumed there, or that the large magnetic field gradients close to the Sun leads to effective particle focusing/mirroring (e.g., Roelof 1969) which is not included in the standard Parker transport equation, making the simulation invalid in the innermost regions of the heliosphere.

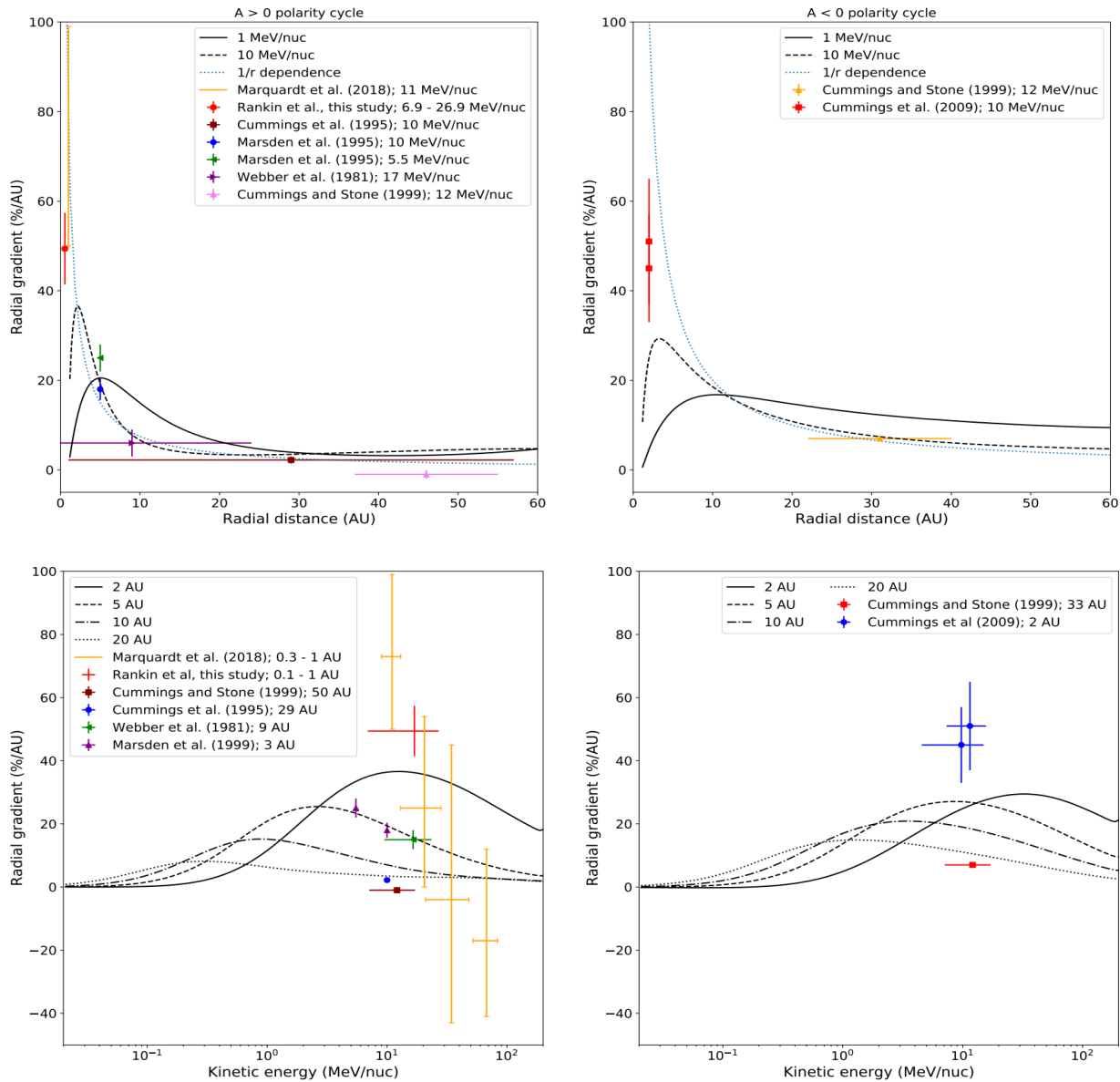

Figure 5. Computed and measured radial gradient as a function of radial distance (top panels) and kinetic energy (bottom panels) for the qA > 0 (left panels) and qA < 0 (left panels) drift cycles. Model results from Strauss & Potgieter (2010) are included, showing the inability of the model to reproduce the measured gradients inside ~2 au.

More generally, the results demonstrate that large radial gradients are persistent close to the Sun (in to 0.1 au) as explored by PSP. Radial gradients observed in the past by previous missions including Pioneer, Voyager and Ulysses were also larger than expected. Standard drift models during qA > 0 solar cycles (e.g., the current one; with drifts downward from the poles) have predicted gradients smaller than those observed (e.g., Jokipii 1987). The assumption made in most of these models is that the heliospheric magnetic field conforms at least generally to the Parker Spiral; the large radial gradients found here suggest that variations in this magnetic structure strongly influences cosmic ray drifts, well inside 1 au; something not captured in the standard ACR transport models.

Previous insights from Ulysses at high latitude may prove critical to guide potential explanations of the persistently large radial cosmic ray gradients during qA > 0 solar cycles. Ulysses observations revealed that the amplitudes of the field variations increase with latitude (Smith et al. 1995) and the latitude gradients are smaller than expected from traditional drift models (McKibben et al. 1995). One argument relates to how the real magnetic field varies with heliocentric distance. The magnetic field for a Parker Spiral winds along cones of constant latitude, and becomes perfectly radial at the pole. Therefore, the field decreases progressively with increasing latitude approaching $r^{-2}$ at the pole. Conversely, the transverse components of field variations decrease as $r^{-1}$ and therefore naturally overwhelm the Parker Spiral component at high latitudes where the cosmic rays access the region near the Sun in cycles with $qA > 0$. These variations in the magnetic field would naturally influence cosmic ray transport to both decrease the latitudinal cosmic ray gradient and increase the radial cosmic ray gradient (Fisk & Schwadron 1995). Such variations – known to occur on intermediate scales (between hundredths of an au and a few au) – should be particularly effective in influencing ACR oxygen owing to its large gyroradius[2] (~0.01 au; ~16 mass per charge).

In the context of PSP measurements, a variety of fluctuations have been observed to affect the field near the Sun. Those that survive will tend to be transverse (e.g., ~$r^{-1}$) and therefore drive the field away from the idealized near-radial Parker Spiral close to the Sun. The impact of tansverse field variations described above affect cosmic ray transport near the Sun, making this a compelling topic for future investigation.

In this study we have presented the first measurements of cosmic rays in to 0.01 au and have found that both the energy spectra and the radial intensity gradient of ACR oxygen show remarkable agreement with prior observations in the inner heliosphere across multiple epochs of solar minima. The ACR component of the spectra is similar across the minima of solar cycles 20/21, 23/24, and 24/25 (other minima will be the subject of future study), consistent with the modern trend at 1 au, while recent GCR spectra (SC 23/24 and 24/25) deviate from the past (SC 20/21). The magnitude of the radial intensity gradient ($49.4 \pm 8.0\%$ au$^{-1}$; 6.7 to 27 MeV/nuc) observed by PSP from ~1.0 to 0.01 au is nearly identical to that measured by Helios ~45 years ago, from 1 to 0.3 au over similar energies ($48 \pm 12\%$ au$^{-1}$; 9 to 29 MeV/nuc) and is larger than predicted by models. A possible explanation is that intermediate scale variations in the magnetic field structure (deviating from that of the typical Parker Spiral) strongly influence cosmic ray transport. We look forward to continued measurements near the Sun by PSP (and Solar Orbiter; Müller et al. 2020), as they will enable unprecedented views of evolving cosmic ray gradients progressively closer to the Sun and through the development of solar cycle 25.

## Acknowledgement

*This work was supported as a part of the Integrated Science Investigation of the Sun on NASA's Parker Solar Probe mission, under contract NNN06AA01C. The ISOIS data and visualization tools are available to the community at https://spacephysics.princeton.edu/missions-instruments/isois; data are also available via the NASA Space Physics Data Facility (https://spdf.gsfc.nasa.gov/). Parker Solar Probe was designed, built, and is now operated by the Johns Hopkins Applied Physics Laboratory as part of NASA's Living with a Star (LWS) program. Support from the LWS management and*

---

[2]The gyrodius of singly charged oxygen at our mean energy (13.4 MeV/nuc) in a ~5 nT field is ~0.01 au, which is roughly equivalent to that of a 1.5 GeV proton and is also of the order of the correlation length of turbulence in the solar wind (also ~0.01 au; Matthaeus et al. 2005; Isaacs et al. 2015). In contrast, the same energy proton has a gyroradius of ~0.0007 au (in the inertial range).

technical team has played a critical role in the success of the Parker Solar Probe mission. We thank all the scientists and engineers who have worked hard to make PSP a successful mission. We also acknowledge the support provided by Michigan State University's National Superconducting Cyclotron Laboratory, Texas A&M University's Cyclotron Institute, and the Lawrence Berkeley National Laboratory's 88-inch Cyclotron Laboratory during EPI-Hi's calibration and testing.

# References

Barghouty, A. F., Jokipii, J. R., & Mewaldt, R. A. 2000, AIPC, 528, 337
Bastian, T. S., McKibben, R. B., Pyle, K. R., & Simpson, J. A. 1981, Proc. ICRC, 10, 88
Bobik, P., Boella, G., Boschini, M. J., et al. 2012, ApJ, 745, 132
Boschini, M. J., Della Torre, S., Gervasi, M., et al. 2020, ApJS, 250, 27
Christian, E. R., Cummings, A. C., & Stone, E. C. 1988, ApJL, 334, L77
Cohen, C. M. S., Christian, E. R., Cummings, A. C., et al. 2021, A&A, Accepted
Cummings, A. C., Stone, E. C., & Webber, W. R. 1987, GRL, 14, 3
Cummings, A. C., Mewaldt, R. A., Stone, E. C., & Webber, W. R. 1990, Proc. ICRC, 6, 206
Cummings, A. C., Mewaldt, R. A., Blake, J. B., et al. 1995, GRL, 22, 4
Cummings, A. C. & Stone, E. C. 1999, AdSpR, 23, 3
Cummings, A. C., Stone, E. C., & Steenberg, C. D. 2002a, ApJ 578, 194
Cummings, A. C., Stone, E. C., & Steenberg, C. D. 2002b, ApJ 581, 1413
Cummings, A. C., Tranquille, C., Marsden, R. G., Mewaldt, R. A., & Stone, E. C., 2009, GRL, 36, L18103
Drake, J. F., Opher, M., Swisdak, M., & Chamoun, J. N. 2010, ApJ, 709, 963
Drews, C., Berger, L., Taut, A., & Wimmer-Schweingruber, R. F. 2016, A&A, 588, A12
Engelbrecht, N. E., Strauss, R. D., le Roux, J. A., & Burger, R. A. 2017, ApJ, 841, 107
Fisk, L. A., & Gloeckler, G. 2009, AdSpR, 43, 1471
Fisk, L. A., & Schwadron, N. A. 1995, JGR, 100, A5, 7865
Fisk, L. A., Kozlovsky, B., & Ramaty, R. 1974, APJL, 190, L35
Fisk, L. A., Wenzel, K. P., Balogh, A. et al. 1998, SSRv, 83, 179
Fox, N. J., Velli, M. C., Bale, S. D., et al. 2016, SSRv, 204, 7
Fu, S., Zhang, X., Zhao, L., & Li, Y. 2021, ApJS, 254, 37
Fujii, Z., & McDonald, F. B. 1997, JGR, 102, A11
Fujii, Z. & McDonald, F. B. 2001, AdSpR, 27, 3
Garcia-Munos, M., Mason, G. M., & Simpson, J. A. 1973, ApJ, 182, L81
Giacalone, J., Drake, J. F., & Jokipii, J. R. 2012, SSRv, 173, 283
Guo, F., Jokipii, J. R., & Kota, J. 2010, ApJ, 725, 128
Guo, Y., Thompson, P., Wirzburger, J., et al. 2021, AcAau, 179, 425
Hajra, R. 2021, SoPh, 296, 33
Hovestadt, D., Vollmer, O., Gloeckler, G., & Fan, C. Y. 1973, PRL, 31, 650
Isaacs, J. J., Tessein, J. A., & Matthaeus, W. H. 2015, JGR, 120, 868
Jokipii, J. R. 1971, RevGSP, 9, 1
Jokipii, J. R. 1987, Natur, 330, 109
Jokipii, J. R. 1996, ApJ, 466, L47
Jokipii, J. R., Levy, E. H., & Hubbard, W. B. 1977, ApJ, 213, 861
Jokipii, J. R., & Thomas, B. 1981, ApJ, 243, 1115
Klecker, B., Mewaldt, R. A., Bieber, J. W., et al. 1998, SSRv 83, 259
Kóta, J., & Jokipii, J. R. 2008, AIP Conf Proc, 1039, 397
Kóta, J. 2010, ApJ, 723, 393
Leske, R. A., Cummings, A. C., Mewaldt, R. A., & Stone, E. C. 2013, SSRv, 176, 253

Marquardt, J., Heber, B., Potgieter, M. S., & Strauss, D. T. 2018, A&A 610, A42

Marsden, R. G., Sanderson, T. R., Tranquille, C., et al. 1999, AdSpR, 23, 3

Matthaeus, W. H., Dasso, S., Weygand, J. M. et al. 2005, PRL 95, 231101

McComas, D. J., & Schwadron, N. A. 2006, GRL, 33, L04102

McComas, D. J., Ebert, R. W., Elliott, H. A., et al. 2008, GRL, 35, L18103

McComas, D. J., Bzowski, M., Fuselier, S. A., et al. 2015, ApJS, 220, 22

McComas, D. J., Alexander, N., Angold, N., et al. 2016, SSRv, 204, 187

McComas, D. J., Rankin, J. S., Schwadron, N. A., & Swaczyna, P. 2019a, ApJ, 884, 145

McComas, D. J., Christian, E. R., Cohen, C. M. S., et al. 2019b, Natur, 576, 223

McDonald, F. B. 1974, Proc. IAUS, 57, 415

McDonald, F. B. 1998, SSRv, 84, 33-50

McDonald, F., Fujii, Z., Ferrando, P., et al. 2001, Proc. ICRC, 10, 3906

McDonald, F. B., Webber, W. R., & Reames, D. V. 2010, GRL, 37, L18101

McKibben, R. B., 1987, RvGeo, 25, 3

McKibben, R. B., 1989, JGR, 94, A12

McKibben, R. B., Connell, J. J., Lapote, C., et al. 1995, The High Latitude Heliosphere, 367 (London: Kluwer Academic Publishers)

Mewaldt, R. A., Selesnick, R. S., Cummings, J. R., Stone, E. C., & von Rosenvinge, T. T. 1996, ApJ, 466, L43

Mewaldt, R. A, Davis, A. J., Lave, K. A., et al. 2010, ApJL, 723, L1

Möbius, E., Hovestadt, D., Klecker, B., et al. 1985, Nature, 318, 426

Moloto, K. D. & Engelbrecht, N. E. 2020, ApJ, 894, 121

Moraal, H. & Stoker, P. H. 2010, JGR, 115, A12109

Müller, D., St. Cyr, O. C., Zouganelis, I., et al. 2020, A&A 642, A1

Ngobeni, M. D., & Potgieter, M. S. 2010, AdSpR, 46, 4

Parker, E. N. 1965, PSS, 13, 9

Potgieter, M. S. 1998, SSRv, 83, 147

Potgieter, M. S. 2013, SSRv, 176, 165

Potgieter, M. S. 2017, AdSpR, 60, 4

Rankin, J. S., McComas, D. J., Leske, R. A., et al. 2021, ApJ 912, 139

Roelof, E. C. 1969, in Lectures in High Energy Astrophysics, ed. H. B. Ögelman & J. R. Wayland, Jr. (Washington, DC: NASA SP-199), 111.

Schwadron, N. A., Lee, M. A., & McComas, D. J. 2008, ApJ, 675, 1584

Shen, Z. N., Qin, G., Zuo, P., & Wei, F. 2019, ApJ, 887, 132

Senanayake, U. K., & Florinski, V. 2013, ApJ, 778, 122

Smith, E. J., Neugebauer, M., Balagh, A., et al. 1995, The High Latitude Heliosphere, 165 (London: Kluwer Academic Publishers)

Smith, E. J., & Balogh, A. 2008, GRL, 35, L22103

Stone, E. C., Frandsen, A. M., Mewaldt, R. A., et al. 1998a, SSRV, 86, 1

Stone, E. C., Cohen, C. M. S., Cook, W. R., et al. 1998b, SSRv 86, 357

Stone, E. C., Cohen, C. M. S., Cook, W. R., et al. 1998c, SSRv 96, 264

Stone, E. C., & Cummings, A. C. 1999, Proc. ICRC, 7, 500

Strauss, R. D., & Potgieter, M. S. 2010, JGR, 115, A12111

Strauss, R. D., & Potgieter, M. S. 2014, SoPh, 289, 3197

Strauss, R. D., Potgieter, M. S., Ferreira, S. E. S., & Hill, M. E. 2010, A&A, 522, A35

Wang, Y. M., Robbecht, R., & Sheeley Jr., N. S., 2009, APJ 707, 1372

Webber, W. R., McDonald, F. B., Trainor, J. H., et al. 1975, Proc. ICRC, 12, 4233

Webber, W. R., McDonald, F. B., & Trainor, J. H. 1977, Proc. ICRC, 3, 233

Webber, W. R., McDonald, F. B., von Rosenvinge, T. T., & Mewaldt, R. A. 1981, Proc. ICRC, 10, 92

Wiedenbeck, M. E., Angold, N. G., Birdwell, B., et al. 2017, Proc. ICRC, 35, 16

Zhao, L. L., Adhikari, L., Zank, G. P., et al. 2018, ApJ 856, 94

## Appendix A.

Table 1 lists radial gradients of ACR oxygen in the inner heliosphere compiled from 5 solar cycles of observations, taken at varying radial distances by many spacecraft, including Pioneers 10 & 11 (P10; P11), Voyagers 1 & 2 (V1; V2), Helios, IMP-7, IMP-8, STEREO, Ulysses, ACE, and Parker Solar Probe (PSP).

| Reference | Time Period | Energy Range (MeV nuc$^{-1}$) | Gradient (% au$^{-1}$) | Radial Distance | Spacecraft | Solar Cycle |
|---|---|---|---|---|---|---|
| Marquardt et al. 2018 | 1975 to 1977 | 9 to 29 | 48 ± 12 | 0.3 to 1 au | Helios | 20/21 qA > 0 |
| Webber et al. 1977 | 1972 to 1976 | 11 to 27 | 25 ± 5 | 1 to 5 au | P10 & 11 | 20/21 qA > 0 |
| | | | 12 ± 4 | 5 to 10 au | | |
| Webber et al. 1981 | 1972 to 1978 | 9.5 to 24 | 15 ± 3 | 1 to 15 au | IMP-7, IMP-8, P10 | 20/21 qA > 0 |
| Cummings et al. 1990* | 1987.0 to ~1987.5 | 7 to 25 | ~48 | 1 to 3 au | IMP-8, V1 & 2, P10 & 11 | 21/22 qA < 0 |
| Marsden et al. 1999 | 1997 to 1999 | ~10 | 25 ± 3 | 1 to 5 au | SOHO; Ulysses | 22/23 qA > 0 |
| Cummings et al. 2009 | 2007 to ~2008.5 | 4.5 to 15 | 48 ± 13 | 1 to 2.9 au | ACE; STEREO A & B; Ulysses | 23/24 qA < 0 |
| | | 7.3 to 16 | 51 ± 14 | | | |
| **This study** | **~2018.7 to ~2021.2** | **6.7 to 27** | **49.4 ± 8.0** | **0.096 to 0.94 au** | **PSP** | **24/25 qA > 0** |

Table 1. ACR Oxygen Radial Gradients. *Inferred.